\documentclass[apjl]{emulateapj}

\usepackage[colorlinks,urlcolor=blue,citecolor=blue,linkcolor=blue]{hyperref}

\usepackage[english]{babel}
\usepackage{graphicx}
\usepackage{verbatim}
\usepackage{amsmath}
\usepackage{subfigure}

\newcommand{\be}{\begin{equation}}
\newcommand{\ee}{\end{equation}}

\slugcomment{Published in ApJ,  2014 Aug 14}
\shortauthors{van Velzen \& Farrar}

\begin{document}

\begin{abstract}
We report an observational estimate of the rate of stellar tidal disruption flares (TDFs) in inactive galaxies, based on a successful search for these events among transients in galaxies using archival SDSS multi-epoch imaging data (Stripe~82). This search yielded 186 nuclear flares in galaxies, of which two are excellent TDF candidates. Because of the systematic nature of the search, the very large number of galaxies, the long time of observation, and the fact that non-TDFs were excluded without resorting to assumptions about TDF characteristics, this study provides an unparalleled opportunity to measure the TDF rate. To compute the rate of optical stellar tidal disruption events, we simulate our entire pipeline to obtain the efficiency of detection. The rate depends on the light curves of TDFs, which are presently still poorly constrained.  Using only the observed part of the SDSS light curves gives a model-independent upper limit to the optical TDF rate, $\dot{N}<2\times 10^{-4}\,{\rm yr}^{-1}{\rm galaxy}^{-1}$ (90\% CL), under the assumption that the SDSS TDFs are representative examples.   We develop three empirical models of the light curves, based on the two SDSS light curves and two more recent and better-sampled Pan-STARRS TDF light curves, leading to our best estimate of the rate: $\dot{N}_{\rm TDF} = (1.5 - 2.0)_{-1.3}^{+2.7} \times 10^{-5}~{\rm yr}^{-1} {\rm galaxy}^{-1}$.  We explore the modeling uncertainties by considering two theoretically motivated light curve models, as well as two different relationships between black hole mass and galaxy luminosity, and two different treatments of the cutoff in the visibility of TDFs at large $M_{\rm BH}$. From this we conclude that these sources of uncertainty are not significantly larger than the statistical ones.   Our results are applicable for galaxies hosting black holes with mass in the range of a few $10^{6}$ to $10^8\,M_\odot$, and translates to a volumetric TDF rate of $(4 - 8) \times 10^{-8\pm0.4}\,{\rm yr}^{-1}{\rm Mpc}^{-3}$, with the statistical uncertainty in the exponent.
\end{abstract}

\title{Measurement of the rate of stellar tidal disruption flares}
\shorttitle{The tidal disruption rate from SDSS data}
\author{Sjoert van Velzen\altaffilmark{1} and Glennys R. Farrar\altaffilmark{2,3}}
\email{s.vanvelzen@astro.ru.nl}
\altaffiltext{1}{Department of Astrophysics/IMAPP, Radboud University, P.O. Box 9010, 6500 GL Nijmegen, The Netherlands}
\altaffiltext{2}{Center for Cosmology and Particle Physics, New York University, NY 10003, USA}
\altaffiltext{3}{Department of Physics, New York University, NY 10003, USA}

\maketitle

\section{Introduction}
Perturbations to the orbit of a star can bring it within a few gravitational radii of the supermassive black hole at the center of its galaxy, where the star will be torn apart in the strong tidal gravity field of the black hole.  The resulting electromagnetic burst can outshine the host galaxy for months to years \citep{Rees90}. The stellar debris is ejected into high-eccentricity orbits, and after a time 
\begin{equation}\label{eq:fb}
t_{\rm fb} \approx 0.11 (M_{\rm BH} / 10^6 M_\odot )^{1/2}~{\rm yr} \quad,
\end{equation} 
roughly half of this gas is expected to return to the pericenter at a rate $\dot{M}_{\rm fb}\propto t^{-5/3}$ \citep{EvansKochanek89,Rees88, Phinney89}. Deviations from this single power law description of the fallback rate are expected at early times, with the exact shape depending on the distribution of internal energy in the star \citep*{Lodato09,Guillochon13}. For non-spinning black holes with a mass of $\lesssim 10^8~M_\odot$ the disruption of a solar-type star typically occurs outside the Schwarzschild radius and thus is visible to observers outside the horizon \citep{Hills75}.  For rapidly spinning black holes, the maximum mass for a visible disruption is higher by a factor of $\approx 5$  \citep{Kesden12}.

Only a small number of (candidate) TDFs are known. They were primarily found by searching for short-lived flares in soft X-ray \citep[e.g.,][]{KomossaBade99,Grupe99, Saxton12}, UV \citep{Gezari09,Gezari12}, or optical surveys \citep{vanVelzen10,Cenko12,Chornock14, Arcavi14}, or by looking for the signal that such a flare could leave in the optical spectrum of a galaxy \citep{Komossa08, Wang12}. The properties of the optical/UV TDFs are roughly consistent with the predicted signature of thermal emission from the stellar debris as it falls back onto the black hole \citep{loebUlmer97, Ulmer99, strubbe_quataert09, Lodato11}. Recently, two candidate TDFs with a transient radio counterpart were discovered in $\gamma$-rays by {\it Swift} \citep{Bloom11, Burrows11, Levan11, Zauderer11,Cenko12b}; these non-thermal flares are best explained by a relativistic outflow that was launched as a result of the disruption, seen in ``blazar mode'' \citep{Bloom11}.

The frequency of stellar capture by supermassive black holes depends on how the orbits of stars evolve.  The rate of flares due to the tidal disruption of stars can thus be used to probe the gravitational potential and phase space disruption of stellar orbits in their host galaxies, which are essentially unconstrained by observations for  $z > 0.01$. Furthermore, it will be interesting to compare the rate of tidal disruptions to the production rate of hypervelocity stars. These unbound stars have been observed in the outer Milky Way halo \citep{Brown05}. Their ages \citep{Brown12b} imply that most of them are likely the result of a three-body interaction of a binary star system and the central supermassive black hole \citep{Hills88}, which ejects one binary partner at high speed. It has been suggested \citep{Gould03, Ginsburg06, Perets09} that the members of the binary that remain bound to Sgr~A* could explain the origin of the S~stars \citep[][]{Eckart96,Ghez05} at the Galactic center. Since orbital diffusion of these stars on tight orbits leads to capture by the supermassive black hole, the disruption of stellar binaries could provide a single framework to explain three different phenomena: hypervelocity stars, the S~star cluster, and TDFs  \citep{Bromley12}. 

The rate of tidal disruptions is also important for understanding the origin of the relativistic TDFs discovered by {\it Swift}. If a large fraction of tidal flares are accompanied by a relativistic jet these events will dominate the transient radio sky, and upcoming radio variability surveys should detect tens to hundreds per year \citep{Frail12,vanVelzen12b}. By comparing the radio and optical TDF rates, we can thus determine the fraction of stellar tidal disruptions that launch jets. Measuring this fraction should provide new insight to tidal disruption jet models \citep{Metzger12, vanVelzen11}, such as testing the prediction that a pre-existing accretion disk is required for the production of tidal disruption jets \citep{Tchekhovskoy13}. A measurement of the rate is also required to test the suggestion that jets from stellar tidal disruptions are the primary source of ultra-high energy cosmic rays \citep{fg08}.

The rate of TDFs not been well constrained by observations until now. \citet{Donley02} conducted a systematic search for large amplitude X-ray outbursts using archival data of the {\it ROSAT} All Sky Survey \citep{Voges99} and recovered the three known X-ray flares from inactive galaxies.  From this, they deduced a rate of $9_{-5}^{+9} \times 10^{-6}$~yr$^{-1}$~galaxy$^{-1}$ (1$\sigma$ uncertainty from Poisson statistics).   Although \citet{Donley02} presented a detailed analysis of the complicated selection effects to estimate the effective survey area, they assumed that all galaxies host equally luminous flares, which is not expected theoretically.   \citet{Gezari08} did not have a systematic procedure for finding the UV flare candidates they identified and therefore they could not determine a flare rate, but those authors concluded that a disruption rate of  $\sim 1 \times 10^{-4} \, {\rm yr}^{-1}$ can reproduce the number of UV flares they found, although a rate of an order of magnitude lower is not ruled out due to the uncertainty on their adopted TDF light curve model.  

In this work, we derive the rate of tidal disruptions from a survey of nuclear flares in galaxies using the Sloan Digital Sky Survey (SDSS). The straightforward selection function of this search allows, for the first time, a study of how the inferred disruption rate depends on the assumed flare light curve. In Section~\ref{sec:searchana}, we give a summary of the SDSS search for nuclear flares and explain how we compute the efficiency of this search. Theoretical background on the interpretation of the TDF rate and the existing models of optical emission from TDFs is given in Section~\ref{sec:theo_back}. In Section~\ref{sec:lightcurves}, we discuss in detail the light curve models that we adopted for our analysis. The results are presented in Section~\ref{sec:results} and discussed in Section~\ref{sec:disc}. 

We adopt the following cosmological parameters: $h=0.72$, $\Omega_{\rm m}=0.3$,  and $\Omega_\Lambda=0.7$.

\section{TDF search and rate determination methodology}\label{sec:searchana}

\subsection{Summary of SDSS nuclear flare search}\label{sec:sum}
Our search for optical TDFs \citep[][]{vanVelzen10} was conducted in SDSS Stripe~82 \citep{sesar07,bramich08,frieman08}, which is part of the seventh data release \citep{Abazajian09}. The stripe consists of about 300 square degrees along the celestial equator; it contains three seasons of about three months long with high cadence observations ($\sim 5$~days), plus six more years with (much) sparser sampling. 

The first step of our TDF search was to select galaxies with a flux increase of 10\% or more, detected at the 7$\sigma$ level using the Petrosian flux of the galaxy as cataloged by SDSS  \citep{blanton01, strauss02, stoughton02}. The Petrosian flux essentially measures the total galaxy flux using a circular aperture with a radius that is independent of redshift and robust against changes in seeing. The catalog-based selection yields $\sim 10^4$ galaxies with flare candidates that were processed by a difference imaging algorithm. Nuclear flares were selected based on the distance between the center of the host and the flare in the difference image ($d<0\farcs2$), yielding 186 transients. 

To obtain a high-quality parent sample of potential TDFs, we applied the following criteria to the flux in the difference image:  $m<22$ for at least three nights in the $u$, $g$, and $r$ filters. After removing galaxies that fall inside the photometric QSO locus and removing galaxies with additional variability, two flares remained; we shall refer to these as TDE1 and TDE2.  Using the \citet{HaringRix04} scaling relation yields an estimate of the black hole mass of $M_{\rm BH} \approx 0.6$ and $ 3 \times 10^7 \,M_\odot$, respectively.  Additional analysis and follow-up observations showed that these flares are best explained as stellar tidal disruption events \citep[][]{vanVelzen10}.

\subsection{Analysis}\label{sec:ana}
The number of detected flares in a variability survey that targets $N_{\rm gal}$ galaxies is given by
\begin{equation}\label{eq:ratefull}
N_{\rm TDF} = \tau \, \sum_i^{N_{\rm gal}} \epsilon_i \dot{N}_i  \quad,
\end{equation}
where $\dot{N}_i$ and $\epsilon_i$ are the flare rate and detection efficiency for the $i$th monitored galaxy, and $\tau$ is the survey time.   For the TDF search in Stripe~82, two TDFs were found so $N_{\rm TDF}=2$.   SDSS monitored Stripe~82 with an adequate cadence for a potential TDF to pass our cuts starting in 2000, so $\tau=7.6~{\rm yr}$.  Finally, the number of galaxies monitored in our search, $N_{\rm gal}=1.5 \times 10^6$, is the number of galaxies that have a photometric redshift and are outside the QSO locus.
  
The rate of TDFs is expected to depend only weakly on black hole mass as long as $M_{\rm BH}<10^8\,M_\odot$.  Above this mass, the rate of visible disruptions is suppressed due to the horizon of the black hole \citep[][]{Hills75}, with a cutoff depending on the black hole spin \citep[][]{Kesden12}. We use two different ways to parameterize the decrease of visible TDFs due to these so-called direct captures: a simple step function at the ``classical'' maximum mass,
\begin{align}\label{eq:BHsplit}
\dot{N_i} = \left\{ 
 \begin{array}{l l}
   \dot{N} &  M_{\rm BH}<10^{8}\, M_\odot \\
   0 &  M_{\rm BH}>10^{8}\, M_\odot \\
 \end{array} 
    \right. ,
\end{align}
and an exponential suppression for $M_{\rm BH}>2\times 10^7\,M_\odot$,
\begin{align}\label{eq:BHsplit_Kesden}
\dot{N}_i = \dot{N} \times \exp[-( M_{\rm BH}/3\times 10^7)^{0.9}],
\end{align}
which fits the analytical results for a black hole spin of $a\approx 0.5$ \citep{Kesden12}. Now Eq. \ref{eq:ratefull} can be rewritten to obtain a galaxy-independent rate:
\begin{equation}\label{eq:rate}
\dot{N} = \frac{N_{\rm TDF}} {N_{\rm gal} \tau \, \epsilon} \quad
\end{equation} 
Here we have defined the mean efficiency 
\begin{equation}\label{eq:eff}
\epsilon \equiv N^{-1}\sum_i^N\epsilon_i \quad. 
\end{equation}
with the sum running over galaxies according to Eq.~\ref{eq:BHsplit} or Eq.~\ref{eq:BHsplit_Kesden}.

Computing the rate of TDFs thus boils down to determining the efficiency.  The result will obviously depend on the flare's luminosity and duration, e.g., a long, bright flare will be above the detection threshold long after the peak and thus is more readily detected with a given set of observations. In the next section, we discuss our approach to measure the detection probability.

\subsection{Pipeline model: detection probabilities}\label{sec:eff}
As discussed in Section~\ref{sec:sum}, our detection pipeline consists of two stages: a series of catalog cuts followed by difference imaging. Here we discuss how we measure the efficiency for each stage. 

The catalog cuts are applied to the Petrosian flux of the galaxy, so computing the probability that a simulated light curve passes these cuts is easy. By construction, a nuclear flare always falls inside the original Petrosian radius of the galaxy and this radius is not changed significantly by the presence of this flare. This implies that the new Petrosian flux should, to good approximation, be given by the original Petrosian flux plus the flare flux. We confirmed this empirically by inserting point sources into the images of 100 different galaxies and measuring the new Petrosian flux. The mean magnitude difference between this newly measured Petrosian flux and the original Petrosian flux plus the inserted flux is $-0.02 \pm 0.05$. This difference is negligible, so trivial arithmetic can be used to determine whether a simulated flare in a given galaxy will pass our catalog cuts (i.e., re-running the entire SDSS pipeline to derive new catalog fluxes from a simulated image is not required).

\begin{figure}
\includegraphics[trim=-6mm 6mm 4mm 56mm, clip, width=.45 \textwidth]{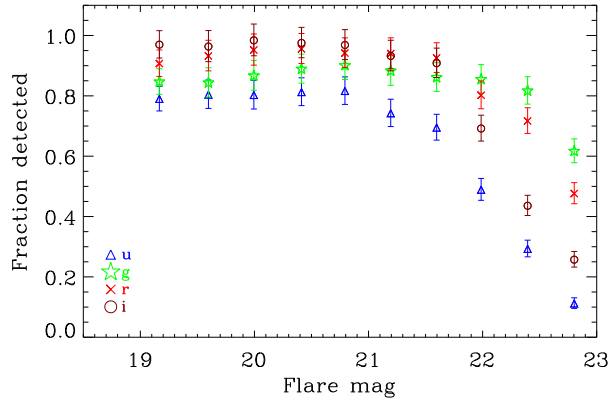}
\caption{Probability of detecting a nuclear flare in the difference image as a function of flare magnitude (each bin contains a range of host galaxy magnitudes). For the TDF search, the flux limit applied to the difference image was $m<22$.}\label{fig:eff_mag}
\end{figure}

Due to variations in seeing on different nights, determining the detection probability in the difference imaging stage of the pipeline is more challenging.  To do so, we selected 1400 galaxies at random and inserted flares at the center of their images.  We then selected three nights per galaxy, drawn uniformly from the set of all observations, and used the point-spread function of each night to create the nuclear flare. Both the host and flare magnitudes were distributed equally in bins between $m=19$ and $m=23$. From the number of detected point sources in each magnitude bin, we obtain the detection probability as a function of both flare and host magnitude. The resulting detection probability as a function of flare magnitude is shown for illustration in Fig.~\ref{fig:eff_mag}. 

To compute the overall efficiency, $\epsilon$, appearing in Eq. \ref{eq:eff}, we first draw a time for the start of the flare from a uniform distribution. We then add the flare flux to the Petrosian flux and check if this galaxy would pass our catalog cuts. In the final step, we use the probability of detection for the given flare and host magnitude to compute whether this flare would be detected at the required level for at least three nights in the $u$, $g$, and $r$ bands. After repeating this process for a large sample of galaxies, the overall efficiency follows from the fraction of flares that are detected by the model pipeline.

Because the flares are inserted into observed galaxy light curves, our method fully takes into account the inhomogeneous cadence and varying data quality of Stripe~82. Our simulation converges after inserting flares into $\sim 10^4$ galaxies; the results presented in Section~\ref{sec:results} are derived using $2\times 10^5$ galaxies chosen at random from the total of $1.2 \times 10^{6}$ galaxies in our analysis.  The numbers of galaxies used for the different stages were large enough to achieve convergence of the result while being computationally efficient.

\section{Theoretical Background}\label{sec:theo_back}
We gather in this section discussions of several theoretical matters pertinent to this study.

\subsection{Tidal disruption rate: theory}\label{sec:theo_rate}
For each location in the galaxy, the set of orbits (in momentum space) that lead to the disruption or capture of a star defines the so-called loss cone. Since this cone empties quickly, theoretical estimates of the TDF rate often boil down to computing the refill rate of the loss cone. The most efficient refill mechanism is the gravitational encounter of stars, which perturbs the orbital angular momentum \citep{Frank_Rees76, Lightman_Shapiro77, Young77}. To quantify the rate of these encounters, the phase space distribution of the stars is needed.  Two different approaches have been used.

Early estimates using nearby galaxies with well-measured surface brightness profiles, \citep{Magorrian_Tremaine99,Syer_Ulmer99} have a scatter of one order of magnitude for black holes of similar mass. More recently, the rate for M32 ($M_{\rm BH}\approx 2\times 10^6$) was estimated to be $\dot{N}=1.7\times 10^{-4}~{\rm yr}^{-1}$ \citep{Wang_Merritt04}.  $N$-body simulations of the diffusion of stars into the loss cone by \citet{Brockamp11} yield a lower rate, and suggest a black hole mass dependence
\begin{align}
\dot{N} = 3.5 \times 10^{-5} \left(\frac{M_{\rm BH}}{10^6 M_\odot}\right)^{+0.31}~ {\rm yr}^{-1} \quad.
\end{align}
(This equation is normalized using the same $M_{\rm BH}$-$\sigma$ relation, Eq.~\ref{eq:Msigma}, that we used to derive Eq.~\ref{eq:ratetheory}.)

Another approach has been to adopt the stellar density profile of a nuclear star cluster, which can be described by an singular isothermal sphere:  $\rho(r) \propto \sigma^2/r^2$, with $\sigma$ the velocity dispersion. For this model, the flux of stars into the loss cone yields the following disruption/feeding rate \citep{Wang_Merritt04}:  
\be
\dot{N} = 7.1 \times 10^{-4} \left(\frac{\sigma}{70\,{\rm km}\,{\rm s}^{-1}}\right)^{7/2}\left(\frac{M_{\rm BH}}{10^6 \,M_\odot}\right)^{-1} ~ {\rm yr}^{-1}.
\ee
Using the empirical relation between black hole mass and velocity dispersion \citep{Ferrarese00,Gebhardt00} as updated in \citet{Graham11},
\begin{equation}\label{eq:Msigma}
\frac{M_{\rm BH}}{10^8 M_\odot} = 1.35 \left(\frac{\sigma}{200\,{\rm km}\,{\rm s}^{-1}}\right)^{5.13}  \quad,
\end{equation}
Eq. \ref{eq:Msigma} leads to 
\begin{equation}
\dot{N} =  9.9\times 10^{-4} \left(\frac{M_{\rm BH}}{10^6\, M_\odot}\right)^{-0.32}~{\rm yr}^{-1} \quad. \label{eq:ratetheory}
\end{equation}
However, while  nuclear star clusters are expected to occur in all low-luminosity stellar spheroids, they can only be resolved for very nearby or large galaxies \citep[e.g,][]{Filippenko03, Ferrarese06} and such galaxies typically host black holes that are too massive to yield visible disruptions. Thus the general applicability of this estimate is not clear. 
 
A further uncertainty arises from various mechanisms that can lead to deviations from the canonical loss cone framework described above. First of all, the galactic potential may be triaxial such that the chaotic orbits of stars bring them close enough to the central black hole to be disrupted even without two-body gravitational encounters \citep{Merritt_Poon04}.   Also, the presence of  a ``massive perturber'', such as a giant molecular cloud, can significantly shorten the relaxation timescale \citep*{Perets07}; see \citet{AlexanderT12} for a review. Finally, the merger of two supermassive black holes is also likely to increase the disruption rate, either simply because the two nuclear star clusters of the two galaxies merge \citep{Wegg_Bode11}, or as a result of the loss cone sweeping through the galaxy due to the recoil of the merged black hole \citep{Komossa08b, Stone11}.  Clearly, a good measurement of the TDF rate can give valuable insight on numerous interesting issues.  

\subsection{Optical emission from TDFs}\label{sec:theo_models}
If an accretion disk forms after the stellar disruption, the luminosity at late time, at a fixed frequency in the Rayleigh--Jeans part of the spectral energy distribution (SED), has the black hole mass dependence and time evolution 
\begin{equation}\label{eq:disk}
L_{\rm TDF} \propto (t-t_D)^{-5/12} \, M_{\rm BH}^{3/4}
\end{equation}
\citep{Lodato11}. The time of disruption ($t_D$) that follows from fitting this power-law decay to the observed light curve is 15, 37 days for TDE1,2. This time is shorter than the typical fallback time (Eq. \ref{eq:fb}), suggesting that this simple disk model does not provide a good description of the early part of the observed  light curve.  Therefore, we must explore more advanced models.

Besides an accretion disk, an important component of optical emission from TDF could be an outflow driven by photon pressure. This wind is expected, since for $M_{\rm BH} \lesssim 5\times 10^7\,M_\odot$ the fallback rate exceeds the Eddington limit \citep[e.g.,][]{Ulmer99}.  Because the temperature of the photosphere of the wind is a function of black hole mass and may increase with time \citep{Strubbe11}, a single power law is not sufficient to describe the optical light curve. One of the light curve models we develop below is based on the disk plus wind emission computed by \citet[][LR11 hereafter]{Lodato11} for the disruption of a star of one solar mass (see their Fig.~3).

A different kind of light curve model is presented in \citet*[][GMR14 hereafter]{Guillochon14}, see also J.~Vinko et al. (in preparation). In this scenario, the luminosity in the UV/optical regime is reprocessed disk emission. Contrary to \citet{Strubbe11} and LR11, the origin of the reprocessing layer is not assumed to be a photon-pressure wind, but is suggested to be due to the ejection of stellar debris. When the accretion rate is super-Eddington, a faction $f_{\rm out}$ of the accretion energy is assumed to be reprocessed by the same layer. The radius of the  photosphere of the reprocessing layer as well as $f_{\rm out}$ are  {\it a priori} unconstrained and need to be obtained by fitting the light curve. Other parameters in the GMR14 model include the mass and polytropic index of the star and the impact parameter. These three parameters yield the fallback rate of stellar debris, as described by the hydrodynamical simulations of \citet{Guillochon13}. As shown in GMR14, this approach yields an excellent fit to the light curve of the well-sampled TDF PS1-10jh \citep{Gezari12}. 

\subsection{Estimating the black hole mass}\label{sec:BHmass}
In the black hole mass regime that is relevant for optical TDFs ($M_{\rm BH}=10^{5-7.5}M_\odot$), few accurate measurements of black hole mass are available \citep[for a review, see][]{Kormendy13}. Most authors assume that the (near) linear scaling of black hole mass with the stellar mass in the bulge \citep[][]{Magorrian98} remains valid in this mass range and we also adopted this approach, using the \citet{HaringRix04} scaling relation. We also consider the conjecture by \citet{Graham12} that the $M_{\rm BH}$-$\sigma$ relation combined with the $L$-$\sigma$ relation yields a broken power law. (This happens because the relation between bulge luminosity and velocity dispersion bends at $M_g\approx -20.5$ \citep{Davies83}, so one should use $M_{\rm BH}\propto L^{2.5}$ for bulge luminosities $M_g > -20.5$, and $M_{\rm BH}\propto L^{1.0}$ otherwise). 

To estimate the bulge magnitude of the galaxies in our sample, we use a method similar to \citet{Marconi04}. The Petrosian flux of the host is multiplied by the bulge-to-total ratio (B/T) determined by \citet{AllerRichstone02} for different Hubble types. Because we are summing over a very large sample, in our simulation it is sufficient to assign the Hubble type of individual galaxies at random, based on the abundance of each type in a flux-limited sample \citep{Fukugita98}. The bulge magnitudes of TDE1,2 are found using the mean B/T for S0 galaxies \citep{AllerRichstone02}. We use the galaxy photometric redshifts of \citet{oyaizu08} to convert between apparent and absolute magnitudes. (Although photometric redshifts are individually subject to error, they are systematically reliable for the large number of galaxies in this study.)

\begin{figure*}
\centering
\subfigure[Phenomenological model.]{
	\includegraphics[trim=5mm 0mm 5mm 10.5mm, clip, width=.44 \textwidth]{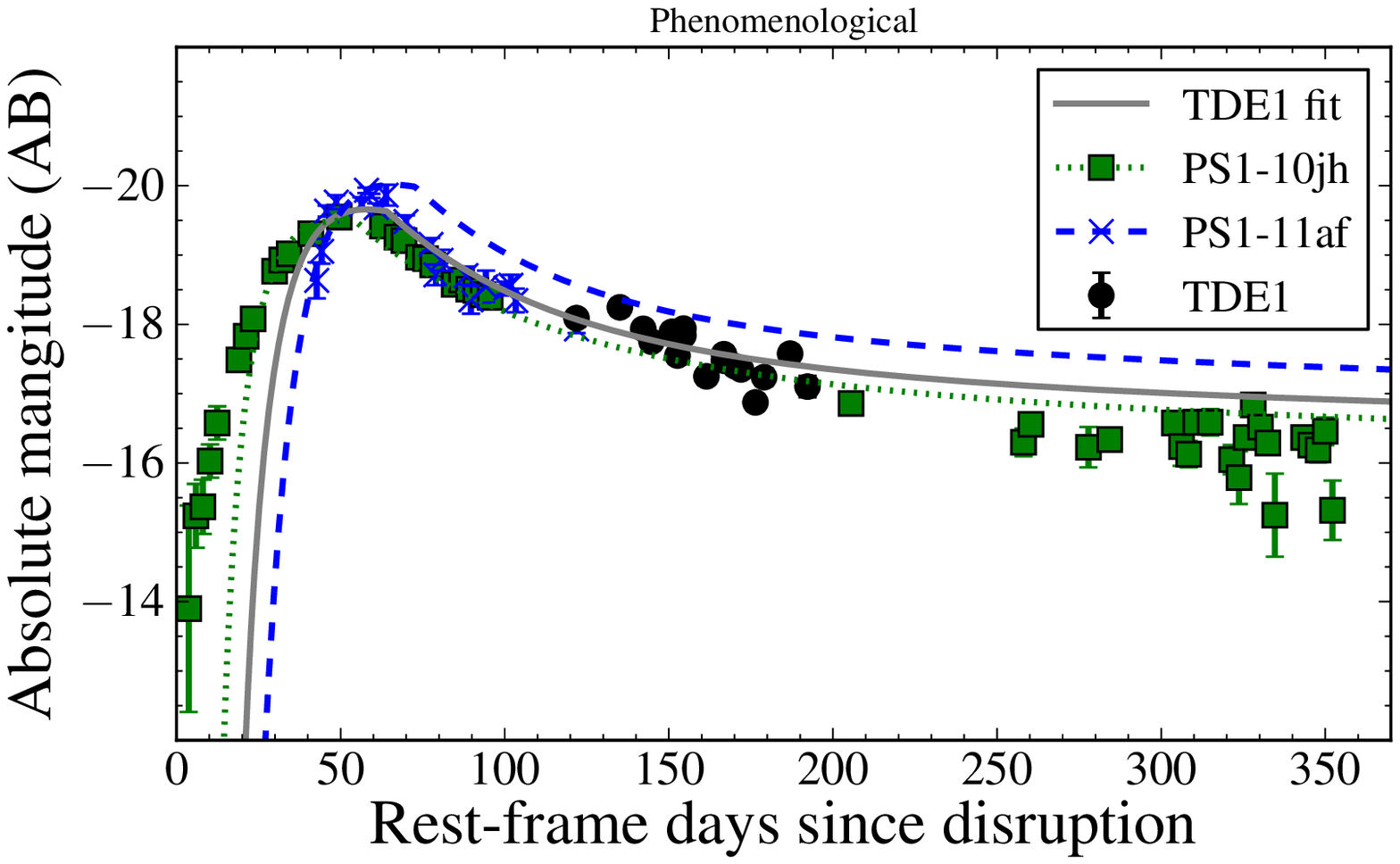}\label{fig:pheno} \quad
	\includegraphics[trim=5mm 0mm 5mm 10.5mm, clip, width=.44 \textwidth]{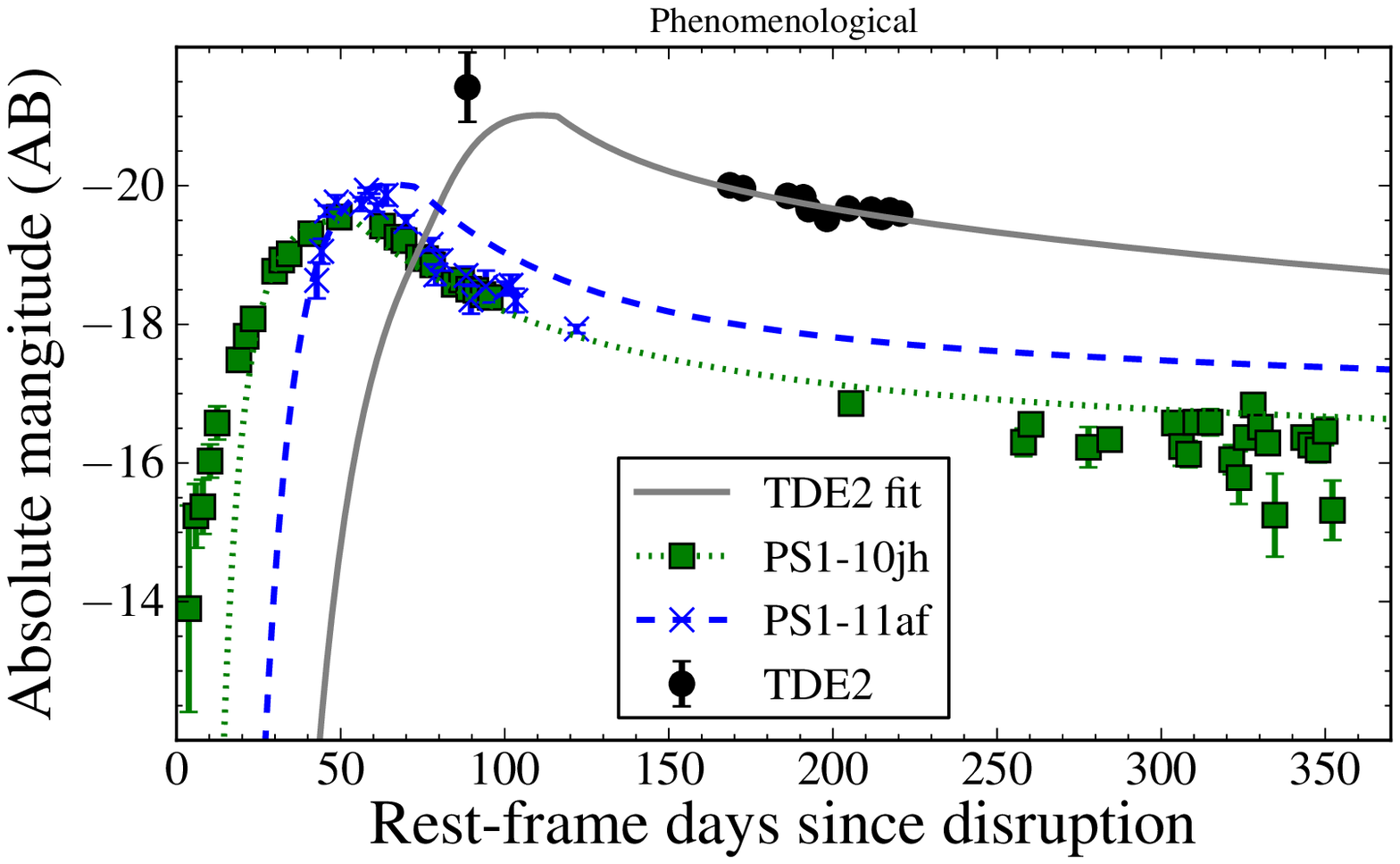}} \\
	\subfigure[Disk + Wind model (based on Lodato \& Rossi 2011).]{
	\includegraphics[trim=5mm 0mm 5mm 10.5mm, clip, width=.44 \textwidth]{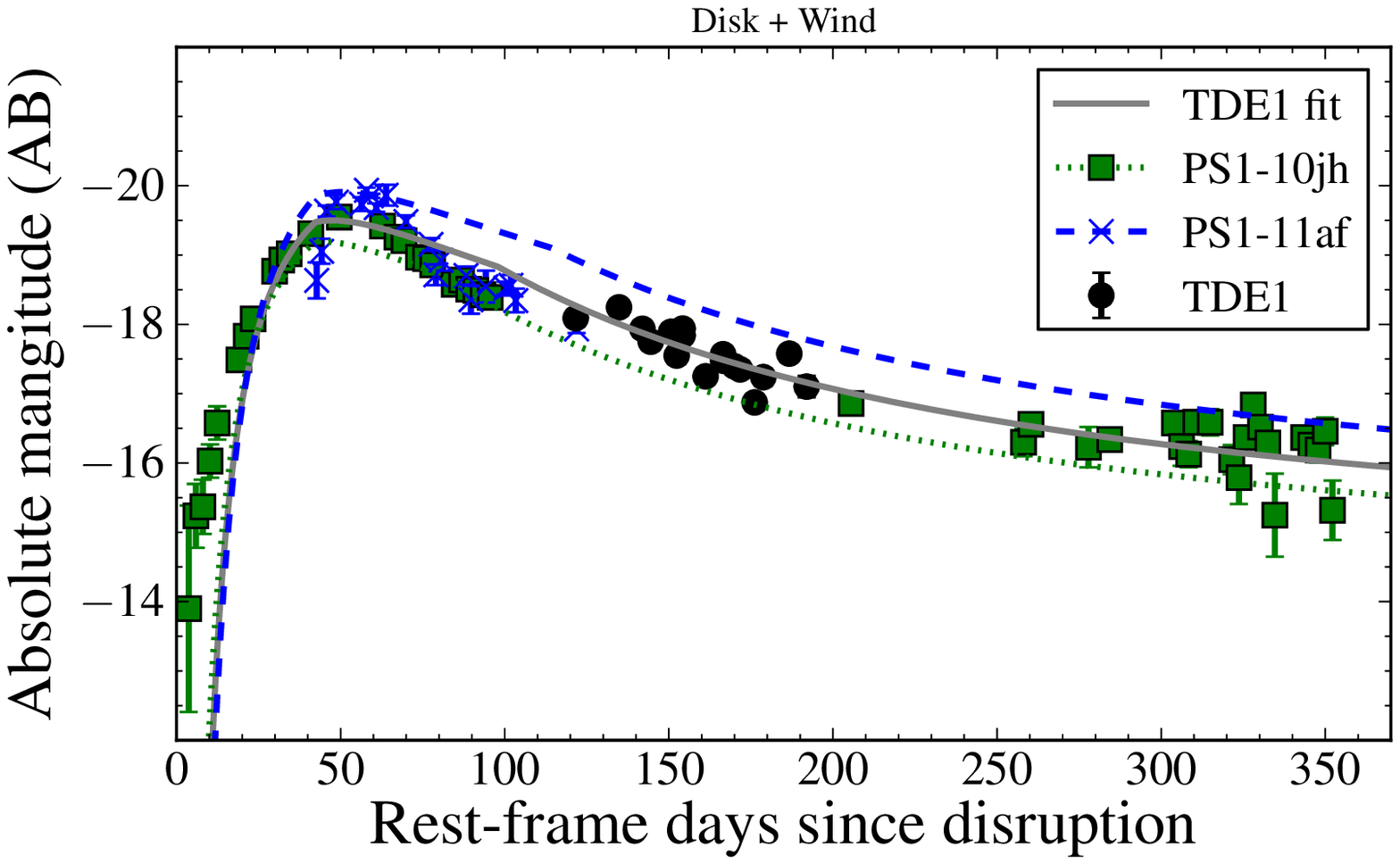}\label{fig:diskwind}  \quad
	\includegraphics[trim=5mm 0mm 5mm 10.5mm, clip, width=.44 \textwidth]{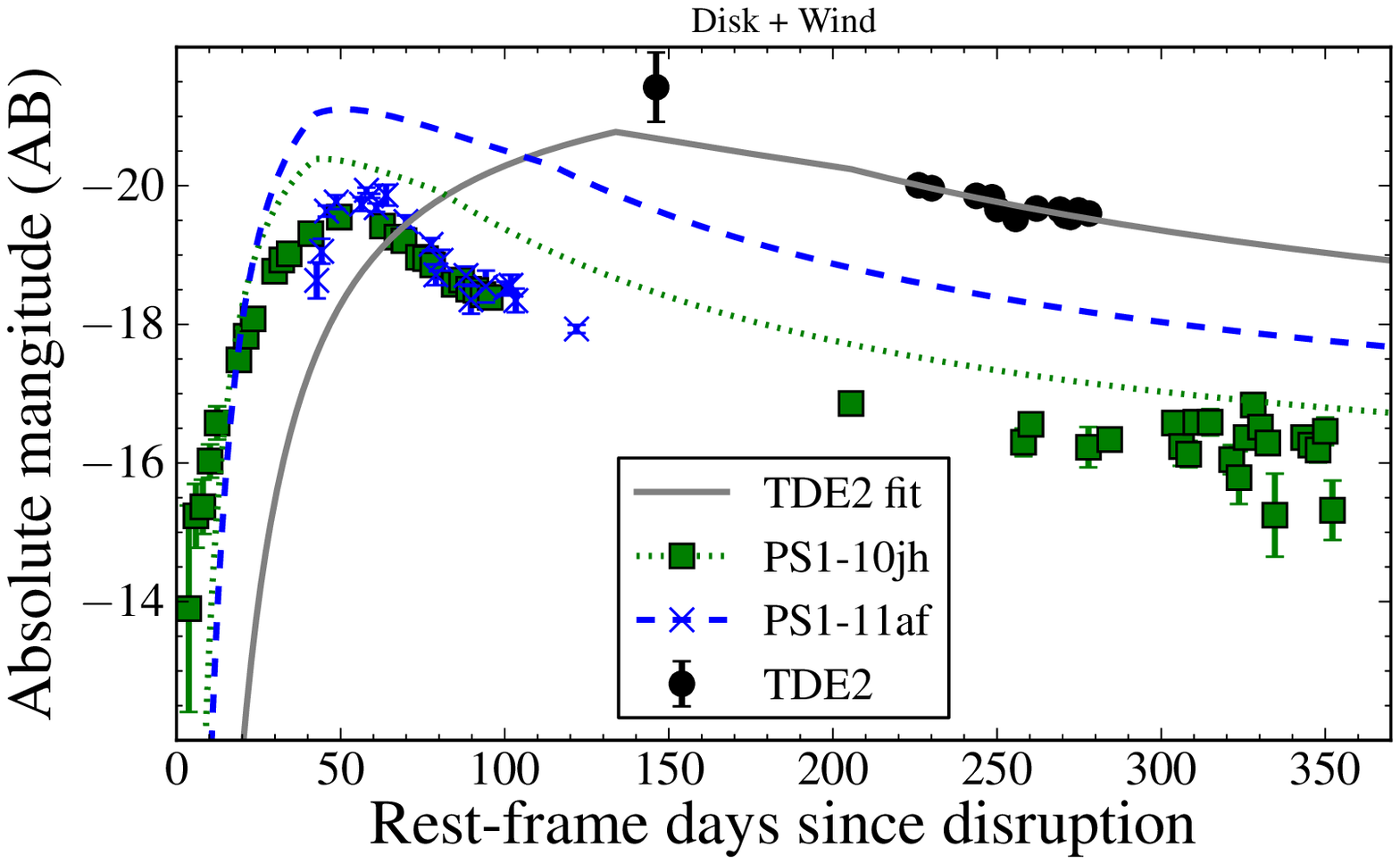}} \\
\subfigure[GMR14 model. ]{
	\includegraphics[trim=5mm 0mm 5mm 10.5mm, clip, width=.44 \textwidth]{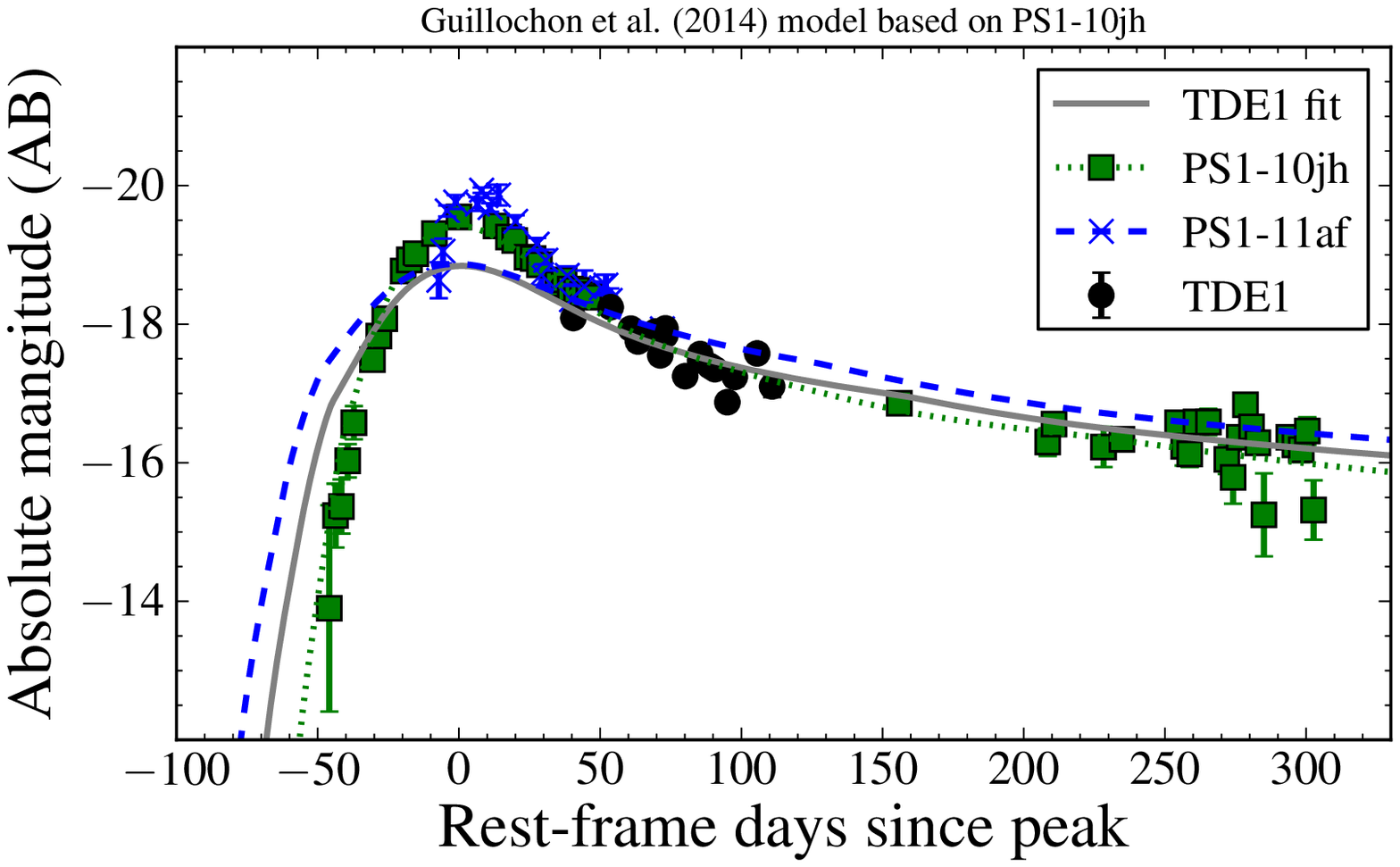}\label{fig:GMR14}  \quad
	\includegraphics[trim=5mm 0mm 5mm 10.5mm, clip, width=.44 \textwidth]{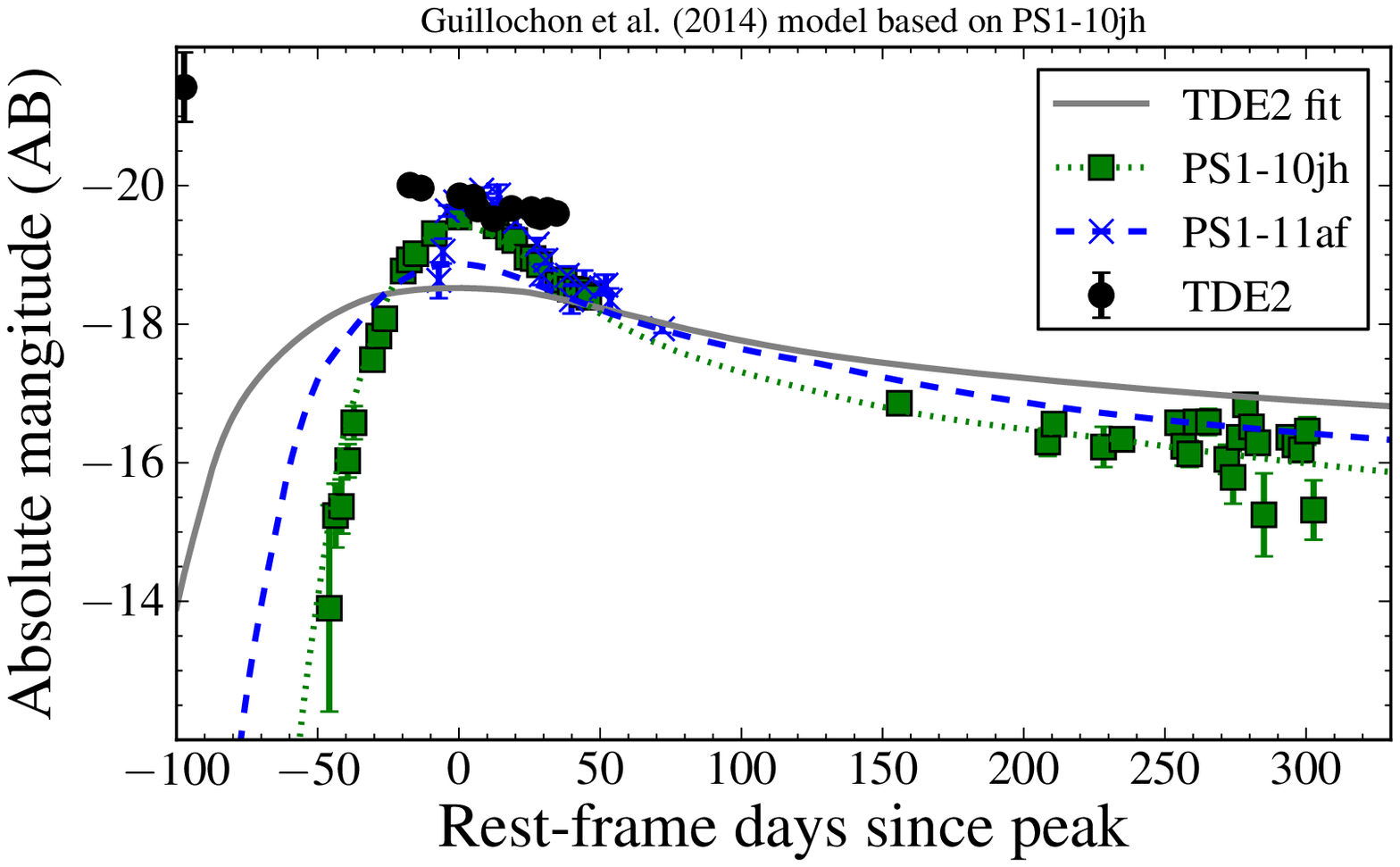}} \\
\caption{Observed $g$-band light curves of TDE1 and 2 (left and right columns, respectively) and PS1-10jh and -11af, compared to  our three different models.  The top and bottom rows show the predictions for the phenomenological and GMR14 models, respectively, in which we fit only for the time of disruption.  In the Disk+Wind model shown in the center row, we allow a different overall normalization for TDE1 and TDE2 as discussed in the text.}\label{fig:lcs}
\end{figure*}

\section{Light Curve Models}\label{sec:lightcurves}
Even with the inherent statistical uncertainty that stems from having only two observed TDFs in our SDSS Stripe~82 search, a major uncertainty of the rate estimate at present is our limited understanding of the optical emission of TDFs.  
The SDSS data for the two TDFs (TDE1 and TDE2, see Section~\ref{sec:sum}) does not completely cover the time they are detectable (i.e., above the flux limit). Their light curves are thus only partially determined so we have to extrapolate the observed light curve forward and backward in time. Furthermore, we also need to know the average light curves of flares from galaxies that host black holes with a mass that is different from the black holes that produced the two SDSS TDFs. We shall approach this problem by using multiple light curve models. Below we present these models, in order of increasing reliance on theory. 

\subsection{Empirical TDF models}\label{sec:empirical_lcs}
Our simplest, ``empirical'' TDF models use no scaling of the luminosity with black hole mass. To keep the results obtained for these models independent of the adopted $M_{\rm BH}$ scaling (see Section~\ref{sec:BHmass}), we do not use a suppression of the rate based on black hole mass (e.g., Eq.~\ref{eq:BHsplit}), but simply compute the per-galaxy rate using only galaxies with a luminosity that is within one magnitude of TDE1 and TDE2. 

\subsubsection{SDSS-only}\label{sec:SDSS-only} We start with a light curve that is identical to the observed light curve: no extrapolation is used. This completely model-independent approach yields an upper limit to the true rate of optical TDF, under the assumption that TDE1 and 2 are reasonably representative. For the computation of the efficiency ($\epsilon$, Eq.~\ref{eq:eff}) we restrict to galaxies with a host luminosity that is within $\pm0.25$~mag of the luminosity of the host of TDE1 ($M_r = -19.9$) or TDE2 ($M_r=-21.3$).

\subsubsection{Pan-STARRS events}\label{sec:PS1events} The two TDFs discovered in Pan-STARRS data, PS1-10jh \citep{Gezari12} and PS1-11af \citep{Chornock14}, have well-sampled light curves, allowing us to use them as example light curves in the computation of the efficiency, under the assumption that all TDFs are similar to the PS1 events.   As shown in Fig.~\ref{fig:lcs}, the light curve of TDE1 is consistent with the post-peak decay rate and luminosity of PS1-10jh, but that of TDE2 is substantially more luminous.   We further note that three TDF candidates that were recently discovered in data from the Palomar Transient Factory \citep[PTF;][]{Law09} by \citet{Arcavi14} have a similar peak luminosity and decay rate as PS1-10jh. This suggests that the two PS1 flares are reasonable examples of true TDFs, for black hole masses similar to those of their host galaxies.  In our simulation of the efficiency using the PS1 light curves, we select either the PS1-10jh or 11af light curve with a probability that is linear with the mass of the host galaxy (e.g., the probability to select the 11af light curve increases from zero at the mass of the host galaxy of 10jh to unity at the mass of its host).  The black-hole masses of the 4 TDEs, TDE1,2, PS1-10jh and -11af, computed as discussed in Section \ref{sec:BHmass}, are respectively $10^{6.8}$, $10^{7.4}$, $10^{6.6}$, and $10^{6.9} \, M_{\odot}$ with about 0.3~dex uncertainty from the scatter in the relation between black hole mass and galaxy luminosity.

\subsubsection{Phenomenological Model}\label{sec:pheno} We also used TDE1 and TDE2 plus the two PS1 TDFs to construct a function that returns a light curve as a function of the mass of the stellar bulge. This `phenomenological model' is simply a collection of power laws that are chosen to roughly reproduce the observed light curves of these four TDFs.  Figure~\ref{fig:pheno} shows the success of this fitting function.  

\subsection{Theory-based models}\label{sec:theoretical_lcs}
The phenomenological model discussed in the previous section provides only a crude way to extrapolate the luminosity of flare with black hole mass. Ideally one would use a framework that yields a set of light curves for a given black hole mass, corresponding to the range of possible disruption parameters (impact parameter, stellar mass, etc.). This is however beyond the scope of this paper because such a framework is not yet available, i.e., it is not yet understood how/where the optical emission of TDFs is produced. 

To further quantify to what extent the uncertainty in TDF light curves impacts our estimate of the TDF rate, we use two different light curve models, which are based on the two models introduced in Section~\ref{sec:theo_models}. 
For both models, we restrict the estimate of the efficiency to galaxies with a bulge luminosity that is with 1~mag of TDE1 or 2. This restriction is imposed to avoid extrapolating the models deep into parameter space that has not been covered by observations. 

\subsubsection{Disk+Wind model}
For the fiducial parameters of LR11, the predicted disk and wind emission is about an order of magnitude lower than the observed luminosity of known TDFs. We therefore renormalized this model such that the total emission (disk plus wind) matches the observed luminosity of TDE1 or TDE2, i.e., we allow a separate normalization of each TDF.  As we remarked above, the decay rate of TDE2 is too steep to fit with only disk emission, but at $M_{\rm BH} >10^7$~$M_{\odot}$, the LR11 model predicts that the disk emission dominates over the emission from the wind. We therefore applied one more modification to the LR11 model, namely multiplying the luminosity of the wind emission with $M_{\rm BH}/5\times 10^6 M_{\odot}$. 

As shown in Fig.~\ref{fig:diskwind}, the resulting `Disk+Wind' model normalized for TDE1 provides a reasonable description of the light curve of PS1~10jh. The Disk+Wind light curve normalized  TDE2 clearly does not reproduce the two PS1 events, which have a lower black hole mass (as estimated from their host galaxy mass) than TDE2. This suggests that the Disk+Wind model parameters obtained for TDE2 should only be used in the highest black hole mass regime of our analysis. We therefore combine the efficiency obtained for TDE1 and TDE2 by weighting the efficiency simulation according to the absolute magnitude of the host galaxy: the probability to select the TDE1-normalized light curve increases from zero at the mass of the host galaxy of TDE2 to unity at the mass of the host of TDE1 (and vice versa).

\subsubsection{GMR14 model}
We can use the model presented in GMR14 to extrapolate the observed light curve of PS1~10jh to galaxies with a lower or higher black hole mass. The \verb TDEfit  software \citep{Guillochon14} was used to obtain the free parameters of that model (the stellar mass, impact parameter, etc), when the black hole mass of PS1~10jh is fixed at the value expected from the \citet{HaringRix04} scaling relation (i.e., the black hole mass was not used as a free parameter in the fit for the parameters of the GMR14 model).  Then, taking those parameters as typical, light curves for other black hole masses are calculated.  As expected, due to its similarity to PS1-10jh ,the GMR14 model provides a good fit for TDE1, whereas the observed light curve of TDE2 exceeds the GMR14 model prediction; see Fig.~\ref{fig:GMR14}. We note TDE2 can be fit within the \citet{Guillochon14} framework, but its best-fit parameters are different from those of PS1-10jh (J.~Vinko et al., in preparation).  

In contrast to the Disk+Wind model, in the GMR14 model, the peak luminosity increases with decreasing black hole mass. We capped the luminosity at the Eddington limit (i.e., $\nu L_\nu<1.3\times 10^{38} M_{\rm BH}/M_\odot\,{\rm erg}\,{\rm s}^{-1}$), which only influences light curves for $M_{\rm BH}<10^{6}\, M_\odot$.  If the flare luminosity does increase to super-Eddington levels at low $M_{\rm BH}$, our rate estimate would not apply to low-mass black holes, and TDFs should be a powerful probe of the low mass black hole population.  

\section{Results}\label{sec:results}
The rate obtained using the suite of light curve models discussed in the previous section is reported below and summarized in Table~\ref{tab:model}. In Fig. \ref{fig:rate} we show the effective-galaxy-years of our pipeline as a function of the bulge luminosity of the host galaxy. The effective-galaxy-years is given by $N_{\rm gal}\times \tau \times  \epsilon$ (i.e., the denominator of Eq. \ref{eq:rate}).

\subsection{TDF rate per galaxy}
Using only the observed SDSS light curve (Section~\ref{sec:SDSS-only}), we find a model-independent upper limit to the rate of optical TDFs: 
\begin{equation}
\dot{N}<2\times 10^{-4} \,{\rm yr}^{-1}{\rm galaxy}^{-1}. 
\end{equation}
Here we used $N_{\rm TDF}<5.3$, the 90\% CL upper limit when two events are detected.  

Using the two TDFs discovered in Pan-STARRS to yield example light curves (Section~\ref{sec:PS1events}) we find
\begin{equation}
\dot{N} = 2.0_{-1.3}^{+2.7}\times 10^{-5}~{\rm yr}^{-1} {\rm galaxy}^{-1}
\end{equation}
 (1$\sigma$ uncertainty for Poisson statistics). Needless to say, this rate is only valid for TDFs that are similar to the two PS1 events.  However given the similarity between these PS1 events and three new TDFs discovered in PTF, it appears to be a reasonable assumption that these light curves are representative of those for black holes with a mass of $\sim 10^{6.5}$. 

\begin{figure}
\centering
\includegraphics[trim=-6mm 6mm 6mm 40mm, clip, width=.45 \textwidth]{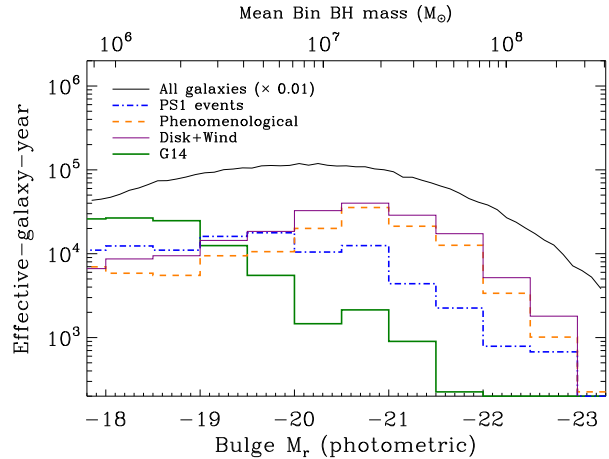}
\caption{Effective-galaxy-years ($N_{\rm gal} \times  \tau \times \epsilon$) for different light curve models in bins of absolute bulge magnitude of the host. We also show the parent galaxy sample by setting $\epsilon=1$ (thin black line). The mean black hole mass in each bin, as obtained using the \citet{HaringRix04} scaling, is indicated on the upper axis.  The bulge luminosities of the hosts of TDE1,2 are $M_r=-19.2, -20.7$. 
}\label{fig:rate}
\end{figure}

For the phenomenological model (Section~\ref{sec:pheno}), we obtain a rate of 
 \begin{equation}
\dot{N} = 1.5_{-1.0}^{+2.0}\times 10^{-5}~{\rm yr}^{-1} {\rm galaxy}^{-1} \quad.
\end{equation}
This rate is slightly lower than the result based on the PS1 events. This happens because the phenomenological model includes TDE2, which increases our estimate of the efficiency for flares from galaxies with $M_{\rm BH} > 10^{7}~M_\odot$, see Fig.~\ref{fig:rate}.

\begin{deluxetable*}{l c c c}
  \tablecolumns{4}
  \tablewidth{0pt}
  \tablecaption{Light curve model efficiencies \& resulting optical TDF rates.\label{tab:model}}
  \tablehead{\colhead{Name} & \colhead{Mean efficiency } &  \multicolumn{2}{c}{TDF Rate }\\ 
   & (\%) & \multicolumn{2}{c}{(${\rm yr}^{-1}{\rm galaxy}^{-1}$)}} \\  
  \startdata
	  SDSS-only & 0.13, 0.62 & \multicolumn{2}{c}{$<1.5 \times 10^{-4}$} \\ [+0.5ex]
	  PS1~events~(10jh,~11af) & 1.0 & \multicolumn{2}{c}{$2.0 \times 10^{-5}$} \\
	  Phenomenological  & 1.4 &\multicolumn{2}{c}{$1.5 \times 10^{-5}$} \\ 
	  \hline \\[-2.0ex]
	  \multicolumn{2}{c}{$M_{\rm BH}$ scaling:} & \multicolumn{2}{c}{Correction for captures:}\\ 
	  \multicolumn{2}{c}{H\"aring \& Rix (2004)}  &\colhead{Step-function}  & \colhead{Exponential} \\ [+0.5ex]
	 \hline \\[-2.0ex]
	  Disk+Wind &0.83, 3.3 & $1.2 \times 10^{-5}$ & $ 1.7 \times 10^{-5}$ \\ 
	  GMR14 &  1.2 & $1.8 \times 10^{-5}$ & $ 1.9 \times 10^{-5}$ \\
	\hline \\[-1.8ex]
	\multicolumn{2}{c}{$M_{\rm BH}$ scaling:} & \multicolumn{2}{c}{Correction for captures:}\\ 
	\multicolumn{2}{c}{Graham (2012)} & \colhead{Step-function} & \colhead{Exponential} \\  [+0.5ex]
	\hline \\[-2.0ex]
	  Disk+Wind & 0.22, 1.5 & $2.1 \times 10^{-5}$ & $3.2 \times 10^{-5}$ \\
	  GMR14 &  1.6 & $1.2 \times 10^{-5}$ & $ 1.3 \times 10^{-5}$
	\enddata
  \tablecomments{In the first column we list the different light curve models. The second column shows the mean efficiency computed using Eq.~\ref{eq:eff}; where the light curve model is based directly on TDE1 and 2, we give the efficiency as obtained for each of them separately.  The tidal disruption rate is shown in the last column(s).  The results shown in the first three rows of this table are independent of black holes mass (Section~\ref{sec:empirical_lcs}). For the two light curve models that depend on black hole mass (Section~\ref{sec:theoretical_lcs}), we compute the rate per galaxy using only those galaxies that can yield visible disruptions. The fraction of visible disruptions is computed in two ways: a step function at $M_{\rm BH}=10^{8}\,M_\odot$ (Eq.~\ref{eq:BHsplit}), and the more realistic exponential suppression due to direct captures (Eq.~\ref{eq:BHsplit_Kesden}).}
\end{deluxetable*}

For the theory-based Disk+Wind and GMR14 models (Section~\ref{sec:theoretical_lcs}), we used two different ways to estimate the suppression of visible TDFs due to the event horizon (shown in the last two columns of Table~\ref{tab:model}), plus two different scaling relations for the black hole mass (shown as a second entry for these models in Table~\ref{tab:model}).  For our Disk+Wind model, the rate increases about 80\% when the \citet{Graham12} scaling is used instead of our default \citep{HaringRix04} scaling relation, while for the GMR14 model the rate decreases by 50\%.  Our two methods of correcting for the event horizon of the black holes yields a 40\%-50\% difference in the derived rate. 
Taking the full range of results gives
 \begin{equation}
\dot{N}_{\rm D+W} = (1.2 - 3.2) \times 10^{-5}~{\rm yr}^{-1} {\rm galaxy}^{-1} \quad
\end{equation}
and 
 \begin{equation}
\dot{N}_{\rm GMR14} = (1.2 - 1.9) \times 10^{-5}~{\rm yr}^{-1} {\rm galaxy}^{-1} \quad.
\end{equation}

The difference between the rate derived for the Disk+Wind and the GMR14 model is relatively small.  This agreement is encouraging since we forced the Disk+Wind model to fit our TDE1 and TDE2, while the GMR14 model was normalized using independent data.  However, we caution that this agreement could be deceptive, since the models predict virtually opposite sensitivity as a function of $M_{\rm BH}$ as evident in Fig. 3.  In the event that GMR14 gives the best description of the flares in the low mass range and the Disk+Wind model is best in the high mass range, the rate estimate would decrease by about a factor of 1.5 with respect to the result based on the phenomenological model.  (The GMR14 model, with parameters tuned to PS1-10jh as used here, does not fit the high mass range (i.e., TDE2) so we do not consider the opposite combination.)

\subsection{Volumetric TDF rate}
To estimate the volumetric rate of TDFs , we compute the efficiency of a given model in bins of galaxy luminosity and integrate this against the galaxy luminosity function ($\phi$). This yields an effective galaxy density: 
\begin{equation}
\rho_{\rm eff} = \frac{\int dM\, \phi(M) \epsilon(M)}  {\int dM  \epsilon(M)} \quad.
\end{equation}
We use the SDSS $r$-band galaxy luminosity function \citep{blanton01}. For integration limits, we adopt $M_r=[-19, -23]$, which covers 90\% of the galaxies in our sample. For a given model, the volumetric rate follows by multiplying the effective galaxy density with the rate per galaxy. 

For the PS1 and the phenomenological models the effective galaxy densities are $4 \times 10^{-3}\,{\rm Mpc}^{-3}$ and $3 \times 10^{-3}\,{\rm Mpc}^{-3}$, respectively.  This corresponds to a volumetric rate of $(4 - 8) \times 10^{-8\pm 0.4}\,{\rm yr}^{-1}{\rm Mpc}^{-3}$ for these two empirical light curve models;  here we have put the statistical uncertainty in the exponent.   For the Disk+Wind model we obtain $\rho_{\rm eff} = 3 \times 10^{-3}\,{\rm Mpc}^{-3}$ (a factor of five lower than the unweighted galaxy density), which implies a volumetric TDF rate in the range $(4 - 10) \times 10^{-8\pm 0.4}\,{\rm yr}^{-1}{\rm Mpc}^{-3}$;  for the GMR14 model, $\rho_{\rm eff}= 5\times 10^{-3}\,{\rm Mpc}^{-3}$. 

\subsection{Comments, Uncertainties and Caveats}
We note that the low value of the mean efficiency of our pipeline, $\epsilon \sim 1\%$, seen in Table~\ref{tab:model}) is a result of defining $\epsilon$ with respect to the full duration of the survey ($\tau=7.6$~yr). Many of the simulated flares are simply not detected because they fall into the gap between two observing seasons or occur in a season with few observations. If we only consider the 3~yr with high cadence observations, the efficiency is a factor of $\sim 10$ higher. 

Our search is most sensitive to galaxies hosting black holes with masses in the range $M_{\rm BH} = (0.5-5) \times 10^{7}\,M_\odot$, as expected for a flux-limited galaxy sample.  The requirement that  $M_{\rm BH}<10^8\,M_\odot$, reduces the galaxy sample by 5\% (or 1\% for the Graham scaling relation), while the correction of direct captures (Eq.~\ref{eq:BHsplit_Kesden}) reduces the sample by 33\% (21\%). Hence the TDF rate for a flux-limited galaxy sample with no restriction on black hole mass can be obtained from Table~\ref{tab:model} using these percentages.  As explained in Section~\ref{sec:sum}, our rate is valid only for galaxies outside the photometric locus of QSO (i.e., our search is not sensitive to TDF inside active galactic nuclei). This cut on the galaxy colors reduced the parent sample by 23\%. 

Finally, we note that obscuration due to circumnuclear dust is a systematic uncertainty in using optical measurements to determine the rate of TDFs.  Some flares will not be detectable at optical frequencies due to extinction, e.g., the (model-dependent) estimate of the extinction for one of the {\it Swift}-discovered TDF (Swift~1644+57) is high, $A_V \sim 3$--5~mag \citep{Bloom11}. The result of extinction by dust is that the optical TDF rate is lower than the intrinsic tidal disruption rate by some factor.  Estimating this factor is non-trivial because the region that obscures the TDF light may occupy only a tiny volume of the full galaxy. The optical spectrum of the host galaxy may therefore not reveal (e.g., via the Balmer decrement) the presence of this dust.   With a larger sample of TDFs, in the future it may be possible to measure the influence of dust via reddening of the TDF SED, depending on the intrinsic variance in the SEDs.

\section{Discussion}\label{sec:disc}
 The optical TDF rate based on our search of SDSS Stripe~82 galaxies is consistent with the rate of large-amplitude, soft X-ray flares from inactive galaxies detected in the {\it ROSAT} All-Sky Survey deduced by \citet{Donley02}, and for most light curve models our rate is within the (very broad) range $ 0.1- 2 \times 10^{-4}\,{\rm yr}^{-1}{\rm galaxy}^{-1}$ of values deemed compatible with the UV observations \citep{Gezari08}.   As noted in the Introduction, the earlier studies were based on more naive treatments of the light curves and dependence on $M_{\rm BH}$ than we have used here and those studies did not attach a systematic uncertainty due to their sensitivity to light curve model.  \citet{Donley02} simply used the median peak luminosity of the X-ray outburst to find the effective volume, ignoring the shape of the light curve and dependence on black hole mass, but did a detailed analysis of their complicated selection effects.  \citet{Gezari08} used a peak luminosity that scaled with black hole mass using the Eddington luminosity fraction function from \citet{Ulmer99}, but used an oversimplified light curve model to estimate their selection function, namely a single blackbody temperature characterized by Eddington luminosity radiation at the tidal disruption radius and a $t^{-5/3}$ power-law decay.

All three studies are hampered by low statistics.  Thus to get a statistically better estimate and to make a proper comparison of the optical TDF rate with the results from {\it ROSAT} and {\it GALEX}, the rates of the latter surveys need to be estimated using the more realistic light curve modeling that we have developed and used here.  With a TDF model covering the entire frequency range of optical, UV, and soft X-rays, the effective-galaxy-years could be determined for {\it ROSAT}, GALEX, and SDSS, and thus derive a rate using the $3+3+2=8$ events discovered by these three surveys.  If this process reveals a significant lack of consistency between the number of events detected in each study individually, it would give useful insight into the validity of the light curve modeling and the importance of systematic effects which will differ from one frequency to another.

Turning now to the comparison with predictions, the analytical disruption rate computed by \citet{Wang_Merritt04} for a singular isothermal sphere (see Eq.~\ref{eq:ratetheory}), $\approx 4 \times 10^{-4}\,{\rm yr}^{-1}$ for our black hole population (see Fig.   \ref{fig:rate}), exceeds our upper limit by a factor of two and our highest estimate of the rate by a factor 10.  In order for the optical TDF rate to be compatible with the \citet{Wang_Merritt04} prediction, either the light curve model must be seriously in error, or if it is accurate, $\approx 90$\% of the flares are obscured in the optical.  (A larger sample of optical TDFs will readily resolve this question because intermediate examples with severe reddening not seen in TDE1 and 2 should show up if most optical TDFs are too obscured to be detected by the pipeline.)  However, not all galaxies may host a nuclear star cluster that can be modeled as an isothermal sphere, and the discrepancy between prediction and our observation may just be a reflection of the breakdown of this hypothesis.  Indeed, the optical flare rate we have determined here is well inside the estimated range of tidal disruption rates for $M_{\rm BH} \sim 10^7\, M_\odot$ based the measured surface brightness profiles of nearby elliptical galaxies, $(1-20) \times 10^{-5}\,{\rm yr}^{-1}$ \citep{Syer_Ulmer99,Wang_Merritt04}.  

\bigskip
\section{Conclusion}\label{sec:finalsum}
We have estimated the rate of TDFs in inactive galaxies implied by the detection of two TDFs in a systematic search for nuclear transients in SDSS Stripe~82 galaxies outside the QSO locus \citep{vanVelzen10} by applying our detection pipeline to simulated light curves. For a given model light curve, the detection efficiency is the fraction of simulated flares that pass all of our selection criteria for the actual cadence and quality of the observations.  The minimal flare model for each event is simply the observed light curve; this yields a model-independent upper limit on the optical TDF rate of $\dot{N}<2 \times 10^{-4}\,{\rm yr}^{-1}{\rm galaxy}^{-1}$ (90\% CL).   

To obtain a more realistic estimate of the TDF  rate, we used a phenomenological model that fits both the TDE1 and TDE2 light curves and light curves from two more recently discovered events in the Pan-STARRS survey, whose sampling covered both the rise and decay of the flares. This gives a rate of
 \begin{align}\label{eq:final}
\dot{N}_{\rm TDF} = (1.5 - 2.0)_{-1.3}^{+2.7} \times 10^{-5}~{\rm yr}^{-1} {\rm galaxy}^{-1} \quad,
\end{align}
with 1$\sigma$ uncertainty for Poisson statistics.  The corresponding volumetric TDF rate is  $(4 - 8) \times 10^{-8\pm 0.4}\,{\rm yr}^{-1}{\rm Mpc}^{-3}$, with the statistical error given as an uncertainty in the exponent.  We also considered two different theoretically motivated light curve models, two different models for how the TDF light curve cuts off at high $M_{\rm BH}$, and two alternatives for the relationship between galaxy luminosity and black hole mass.  From the range of the resultant rates, one can conservatively estimate that the theoretical (i.e., not statistical) uncertainty in $\dot{N}_{\rm TDF} $ is not significantly greater than the statistical one indicated in Eq.~\ref{eq:final}.   

\acknowledgments
We are grateful to E.\,M. Rossi, G. Lodato, and J. Guillochon for supplying their model light curves in tabulated form, and to J. Guillochon for extensive discussions.  We also thank S. Gezari and D. Zaritsky for providing feedback to an early draft of this paper, and W. R. Brown, A. Gal-Yam, M. Kesden, D. Merritt, E. O. Ofek and E. Ramirez-Ruiz for useful discussions. The research of GRF was supported in part by NSF-PHY-1212538.

\bibliography{general_desk}
\end{document}